\documentclass{article}

\usepackage[english]{babel}
\usepackage{CJKutf8}
\usepackage[letterpaper,top=2cm,bottom=2cm,left=3cm,right=3cm,marginparwidth=1.75cm]{geometry}

\usepackage{amsmath, amsthm, amssymb, graphicx}
\usepackage{graphicx}
\usepackage{txfonts}
\usepackage[colorlinks=true, allcolors=blue]{hyperref}

\title{The Whole Pathological Slide Classification via Weakly Supervised Learning}
\author{Qiehe Sun}

\begin{document}
\begin{CJK}{UTF8}{gbsn}
\maketitle

\begin{abstract}
  Due to its superior efficiency in utilizing annotations and addressing gigapixel-sized images, multiple instance learning (MIL) has shown great promise as a framework for whole slide image (WSI) classification in digital pathology diagnosis. However, existing methods tend to focus on advanced aggregators with different structures, often overlooking the intrinsic features of H\&E pathological slides. To address this limitation, we introduced two pathological priors: nuclear heterogeneity of diseased cells and spatial correlation of pathological tiles. Leveraging the former, we proposed a data augmentation method that utilizes stain separation during extractor training via a contrastive learning strategy to obtain instance-level representations. We then described the spatial relationships between the tiles using an adjacency matrix. By integrating these two views, we designed a multi-instance framework for analyzing H\&E-stained tissue images based on pathological inductive bias, encompassing feature extraction, filtering, and aggregation. Extensive experiments on the Camelyon16 breast dataset and TCGA-NSCLC Lung dataset demonstrate that our proposed framework can effectively handle tasks related to cancer detection and differentiation of subtypes, outperforming state-of-the-art medical image classification methods based on MIL. The code will be released later.
\end{abstract}

\section{Introduction}

Histopathological slide examination is widely regarded as the most reliable and accurate standard for clinical diagnosis of many diseases \cite{aeffner2017gold}. However, during the actual diagnostic process, pathologists are required to locate the region of interest (ROI) within the low magnification field of view. They must then carefully examine at the high magnification for signs of tissue structure abnormalities, the presence of a notable number of inflammatory cells, and other relevant factors. In clinical practice, despite the fact that the majority of breast, colon, cervical tissue samples obtained through population screening, as well as numerous lymph node sections removed during surgery from patients, are negative, they still require meticulous screening \cite{van2021deep}. This process can be time-consuming and labor-intensive. To make matters worse, in some regions with limited medical resources, even obtaining a simple pathological report can be challenging, leading to delayed treatment of disease. The situation remained unresolved until the advent of scanners capable of scanning stained pathology sections into pyramid-structured images, known as the whole slide image (WSI), along with the development of artificial intelligence.

\begin{figure}[t]
\centering
   \includegraphics[width=\linewidth,trim=100 0 100 0,clip]{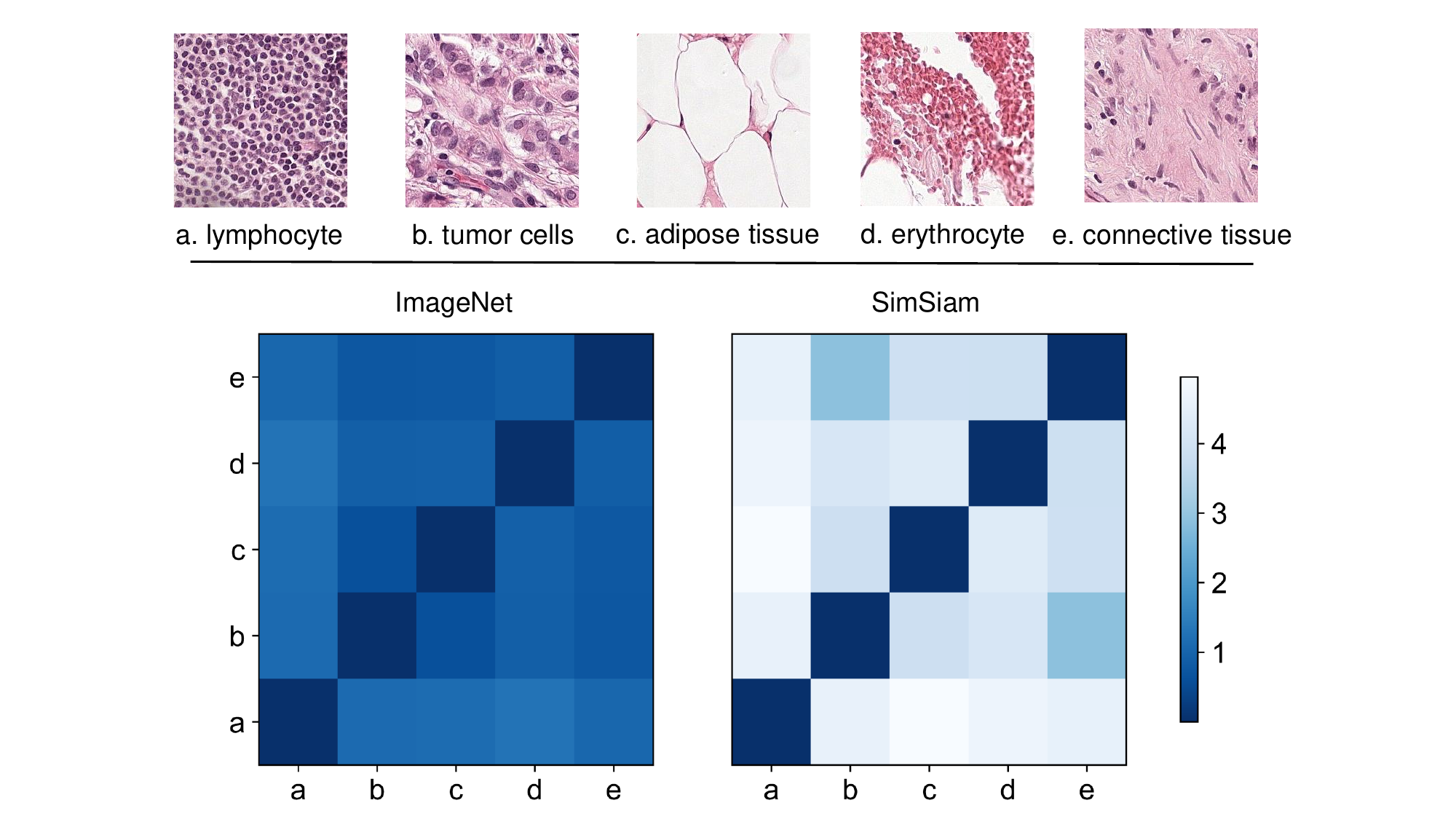}
   \caption{Similarity matrix for five different tissue images from Camelyon16 dataset. We encoded them into 1024-dimensional embeddings using ResNet50 pre-trained on Imagenet (\textbf{Left}), and the one pre-trained by SimSiam, which utilized our proposed data augmentation method (\textbf{Right}), respectively. Euclidean distance is used to describe the similarity between every two feature vectors. To aid in visualization, we applied natural logarithm to the results.}
\label{fig:long}
\label{fig:onecol}
\end{figure}

Due to the immense success of deep learning in natural image tasks, computational pathology has also experienced a significant boost in development. Nevertheless, there are still two major challenges in transferring deep models to the field of pathological images. Firstly, WSI at the highest magnification level is a three-dimensional and high-resolution image that contains at least a billion pixels. Scaling it down to a size that can be processed by GPUs will result in the loss of cellular-level and tissue-level information. The current solution is to cut it into patches with only $10^4$ pixels. However, this approach poses a second challenge ---- obtaining patch-wise labels is difficult and requires experts to annotate millions of images. Slide-level annotations, which are more accessible, only include basic clinical information such as disease progression, molecular subtypes, and survival rates. Therefore, a current research hotspot is how to fully utilize these clinical-level labels without requiring additional manual annotations.

Multiple Instance Learning (MIL) is a special type of weakly-supervised method\cite{dietterich1997solving, quellec2017multiple} that infers fine-grained information through coarse-grained annotations such as clinical diagnoses. In this context, \textbf{slide} and \textbf{patch} correspond to the concepts of \textbf{bag} and \textbf{instance}, respectively, where the attributes of a bag are the sum of the features possessed by its instances. In other words, a positive bag must contain at least one positive instance, while all instances in a negative bag should be negative. The process of MIL involves the extraction, selection, and aggregation of instance features. Various attention-based aggregators constructed by neural networks have been the key to its success in pathological tasks\cite{campanella2019clinical, ilse2018attention, lu2021data, li2021dual, shao2021transmil, thandiackal2022differentiable}, but little research has been done on feature extractors and selection strategies\cite{li2021dual, zhu2022murcl}. 

Most MIL methods use deep residual network pre-trained on the ImageNet\cite{deng2009imagenet} dataset as instance feature extractors\cite{ilse2018attention, lu2021data, shao2021transmil}. However, the texture and color of natural images differ significantly from those of pathological images stained with hematoxylin-eosin (H\&E) dye. To obtain suitable pathological representations without introducing additional supervised signals, self-supervised methods have become crucial. As shown in Figure 1, contrastive learning (CL) methods can effectively distinguish pathological images in feature space, while ResNet\cite{he2016deep} pre-trained on Imagenet fails. 

\begin{figure}[t]
\centering
   \includegraphics[width=\linewidth,trim=100 0 150 50,clip]{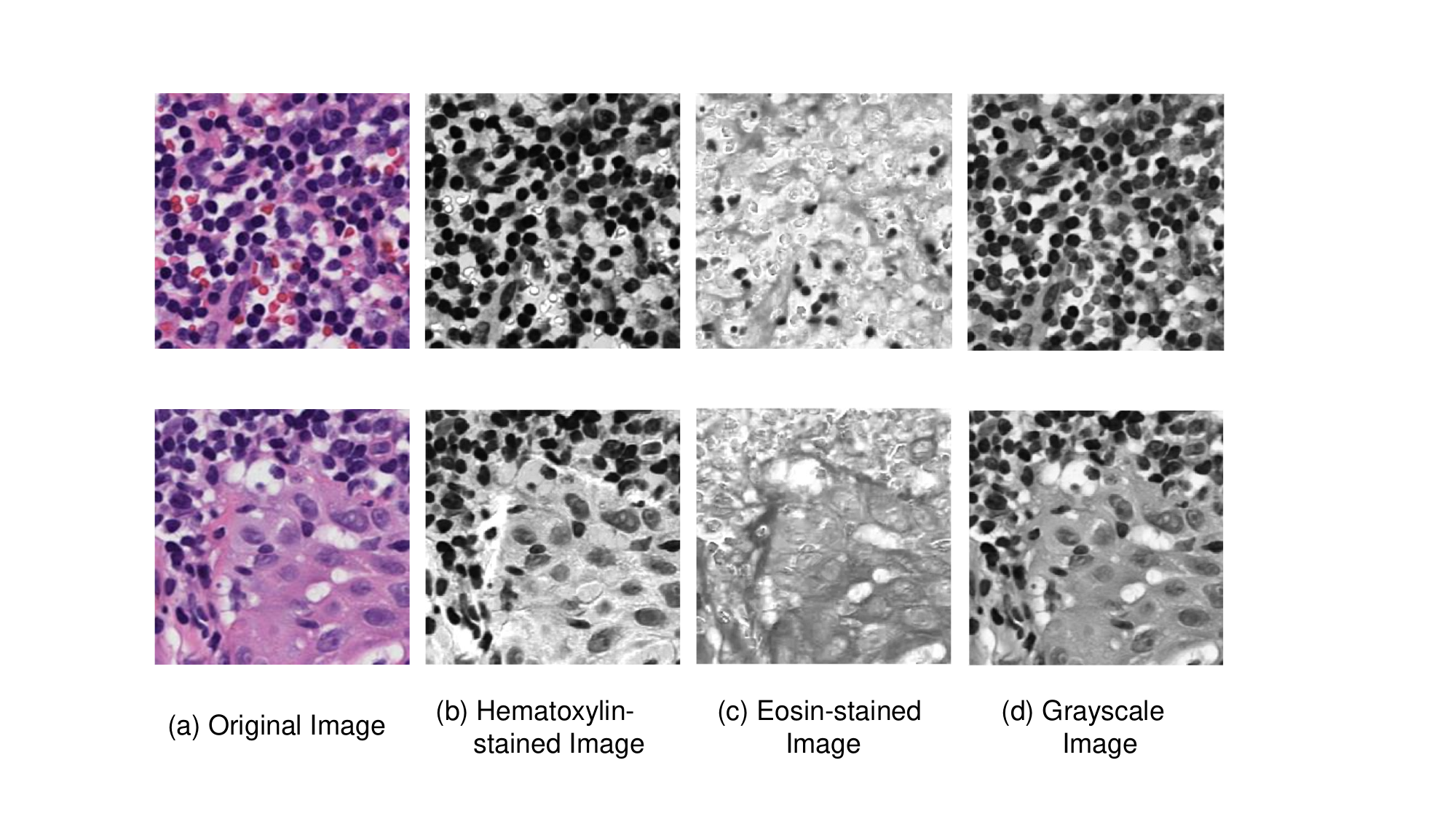}
   \caption{The seperation results. (a) Original H\&E images: separate images of lymphocytes and tumor cells. (b)(c) Hematoxylin and eosin stained images obtained by matrix decomposition of the original image. (d) Grayscale Image.}
\label{fig:long}
\label{fig:onecol}
\end{figure}

Despite some previous works attempted to address this unrealistic situation, they still lack the ability to utilize the inherent inductive biases of pathological images to guide the classification results. In this paper, we introduce a data augmentation method based on stain separation, which is integrated into the existing contrastive learning framework, allowing the feature extractor to focus on more diagnostically valuable information. Stain separation is the process of separating H\&E images into individual images stained with hematoxylin and eosin, as shown in Figure 2. Hematoxylin is a bluish-purple basophilic dye and mainly stains chromatin in the nucleus and nucleic, while eosin is an acidic dye that stains components in the cytoplasm and extracellular matrix red-pink\cite{vahadane2016structure}. Nuclear abnormality is one of the indicators for pathological diagnosis, and the significance of using separated images that are distinguished from grayscale images as sample pairs in contrastive learning lies in separating the foreground of cell nuclei from the background of cytoplasm, thereby guiding the feature encoder to focus more on nuclear variations. In addition, we also introduced another pathological prior: spatial correlation, which means that adjacent patches in spatial position on the WSI have mutual attention. Therefore, we represented the spatial relationship of all tiles in a slide as an adjacency matrix and used it as the input to a graph attention network (GAT) to constrain the attention flow between representations. Based on this consideration, we designed a aggregation network and conducted experiments on two publicly available pathological datasets——Camelyon16 and TCGA-NSCLC, achieving better performance than the state-of-the-art MIL methods. Our main contributions can be summarized as follows:

\begin{itemize}
	\item We proposed a stain-separation based data augmentation technique and applied it to train MIL feature extractors with contrastive learning.
	\item We introduced the absolute positional relationship between tiles to constrain the mutual attention, and designed a graph attention aggregator according to it.
	\item Abundant experiments on two public datasets demonstrate the effectiveness of our framework for slide-level diagnosis.
\end{itemize}

\section{Related Work}
In recent years, the development of deep learning has led to the gradual replacement of MIL algorithms based on shallow structures. We will introduce the current situation of deep MIL models from two perspectives: their development and applications.

\subsection{Deep Multiple Instance Learning}
Early frameworks used simple maximum or average pooling as feature aggregators\cite{feng2017deep, pinheiro2015image}, but subsequent studies suggested that parameterized neural networks were better suited for fitting the contributions of different instances and achieving better results\cite{shao2021transmil, feng2017deep, wang2018revisiting}.  Ilse et al.\cite{ilse2018attention} categorized deep MIL methods into embedding-level and instance-level, following the theorem proposed by Zaheer et al.\cite{zaheer2017deep}. The key component of embedding-level approach is the aggregator, which incorporates attention mechanisms to account for the varying contributions of individual instances to the bag\cite{ilse2018attention, lu2021data}. The advent of Transformer\cite{vaswani2017attention} has enabled the use of self-attention mechanisms for modeling intrinsic relationships between instances, and it has been demonstrated to reduce the information entropy of MIL, thereby mitigating uncertainty\cite{shao2021transmil}.

It should be noted that some recent works have recognized the importance of effective bag embeddings. Due to learnability, instance-level approach is employed for training the extractor\cite{campanella2019clinical}. Nevertheless, such approaches are inherently designed for binary classification problems and may not be well-suited for other types of tasks. To obtain more universal feature extractors, contrastive learning methods, such as SimCLR\cite{chen2020simple} and DINO\cite{caron2021emerging}, have been applied to maximize the separation of patches in the feature space\cite{li2021dual, chen2022scaling}. Aside from contrastive learning, variational autoencoders (VAEs) and generative adversarial networks (GANs) can also be used as methods for training feature extractors\cite{tellez2019neural}. Several studies have also addressed how to filter instance-level embeddings to obtain the optimal bag-level representations, and reinforcement learning (RL) has been employed to select the most representative patches instead of random selection\cite{zhu2022murcl}.

As WSIs are organized in a pyramid data structure, graph neural networks (GNNs), eg. graph convolutional neural networks (GCNs) have been utilized to model the inter-layer and intra-layer relationships\cite{chen2021diagnose, hou2022h, zhao2020predicting}. Although we employed graph neural network like the aforementioned works, our purpose was to constrain inter-instance attention and learn more interpretable bag representations. To integrate the multi-scale information inherent in pathological images, patch-level features at different magnifications have been utilized as input to the aggregator to model both fine-grained and coarse-grained information of the diseased tissue\cite{li2021dual, thandiackal2022differentiable, chen2022scaling}.

\subsection{Pathology Applications based on MIL}
In the analysis of whole slide images, multiple instance learning is widely used due to its label-efficiency and interpretability. This has been demonstrated on several large, diverse, private datasets, including but not limited to colorectal cancer, lung cancer, prostate cancer, bladder cancer, and skin cancer\cite{xu2014weakly, hou2016patch, campanella2019clinical, yao2020whole}. However, in practice, MIL models have not always met the clinical requirements for small datasets. To address this issue, Zhang et al.\cite{zhang2022dtfd} introduced the concept of "pseudo-bags" to artificially expand the dataset. MIL has also shown excellent performance in segmentation, clustering, and other tasks\cite{xu2014weakly, xu2019camel}. Moreover, the success of MIL on immunohistochemistry (IHC) images has opened up possibilities for multimodal analysis beyond just hematoxylin-eosin (H\&E) images\cite{hou2022h}. 

\section{Method}
We developed a weakly-supervised learning framework for slide-level classification based on two pathological priors. In this section, we will describe how we incorporated these priors into MIL framework and provide an overview of our model.

\subsection{Multiple Instance Learning}
Datasets used for multiple instance learning typically contain instances and bags with a hierarchical relationship. There are $N$ bags $\left \{ \left ( X_{i}, Y_{i} \right ) \right \} ^N_{i=1}$ in a dataset. Each bag consists of $n$ instances, where $n$ varies across different bags. Then the label of $X_{i} =  \{ x_{i}^{1} , x_{i}^{2},\cdots, x_{i}^{n} \mid  x_{i}^{k} \in \mathbb{R}^{H \times W \times C_h} \}$ is $Y_{i}\in \mathbb{R}^C$. In the above equations, $C_{h}=3$ for RGB images while $C$ denotes classes for classification task. Assuming that the true label of an instance is denoted by $y_{i}^{k} \in \mathbb{R}^{C}$ which is actually unknown,  in binary problem, we define MIL as: 

\begin{align}
    Y_i=
    \begin{cases}  0, & \text{ iff } \sum_{k=1}^{n}y_{i}^k=0  \\  1, & \text{ otherwise}
    \end{cases}
\end{align}

\noindent Expanding the above equation to multiclass classification, we have:
\begin{align}
    Y_i = S(X_i)=g(\sum_{x_i\in X_i}{f(x_i)})
\end{align}
\noindent Where $S\left ( \cdot \right )$ is a scoring function for instances in bag $X_i$, which is permutation-invariant to $x_i$, while $f\left ( \cdot \right )$ and $g\left ( \cdot \right )$ are two different transformations\cite{ilse2018attention}. Depending on the choice of transformations, MIL can be classified into two categories: instance-level approach and embedding-level approach. 

 Instance-level approach utilizes an instance-level classifier as $f\left ( \cdot \right )$, with $g\left ( \cdot \right )$ being an identity function. However, insufficient training during training may introduce unnecessary error. On the other hand, embedding-level approach tends to construct a bag-level aggregator as $g\left ( \cdot \right )$, with $f\left ( \cdot \right )$ serving as a feature extraction network that is solely used to generate instance embeddings. Nonetheless, due to the absence of large-scale pathological databases such as ImageNet, the extractor may fail to accurately capture the crucial features of instances.  We rethought the fomulation of MIL and employed contrastive learning to train the feature extractor on top of the embedding-level methods. By using such a self-supervised method, we can guide the representations of slides with specific data augmentation techniques while minimizing the initial error.
 
%-------------------------------------------------------------------------
\subsection{Two Priors}
\noindent\textbf{Nuclear Heterogeneity of Diseased Cell.} \quad Abnormalities in the nucleus and chromosomal organization are hallmarks of many diseases, including cancer\cite{uhler2018nuclear}. Pathologists rely on these aberrations to diagnose and grade tumors. For instance, in low-grade ductal carcinoma in situ (DCIS) of the breast, cells are small, regular, and evenly distributed, with nuclei located centrally. By contrast, high-grade carcinoma features large and irregular nuclei. Intermediate-grade falls somewhere in between. The degree of malignancy is directly associated with the tumor's rate of progression, metastasis, and patient survival\cite{hayward2020derivation}. Hematoxylin and eosin (H\&E) is one of the most widely used staining methods in pathological diagnosis. Hematoxylin displays a high affinity for chromatin within the cell nuclei, yielding a bluish-purple hue of the cell nucleus. Then the presence of nuclear abnormalities can be assessed by pathologists through visual observation. Motivated by this, we decomposed H\&E-stained RGB images into H and E components, and subsequently designed an image data augmentation technique to guide the encoder in sensitively capturing the morphological changes of the cell nucleus. In the subsequent section, we will provide a detailed exposition of this data augmentation method.

\vspace{2ex}
\noindent\textbf{Spatial Correlation of Pathological Tiles.} \quad Almost all computational pathology methods involve dividing WSIs into patches to accommodate GPU memory. In the context of MIL, these patches are regarded as individual instances, and their collective representation forms the basis for evaluating the corresponding WSI. Given that the cutting process is typically automated and uncontrollable, a region of cancerous tissue may be shared among several patches, leading to spatially adjacent patches having similar properties. Consequently, during the aggregation process, graph attention may be more appropriate than self-attention, as non-adjacent patches, despite belonging to the same class, lack inherent coupling.

\subsection{Pathological Prior Based MIL}

\begin{figure*}[t]
\centering
   \includegraphics[width=\linewidth,trim=120 30 150 20,clip]{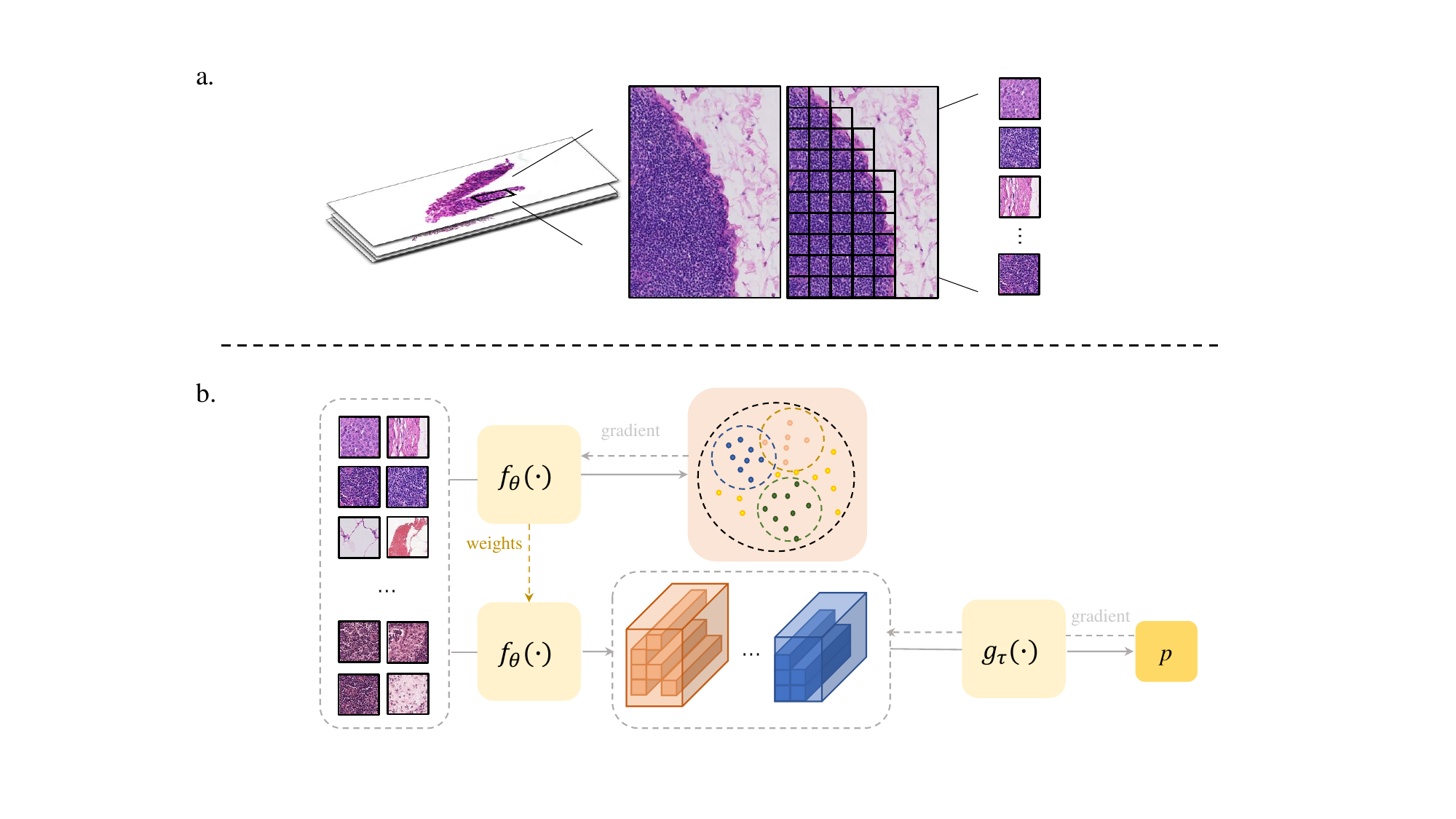}
   \caption{Overview of our architecture. (a) We begin by separating foreground from background by OTSU algorithm and then slicing the whole slide images (WSIs) into patches at a specific magnification. (b)The training is divided into two stages. In the first stage, we employ SimSiam to train the feature extractor $f_{\theta } \left ( \cdot \right )$. During training, we apply data augmentation based on H\&E stain separation to introduce pathological prior, which enables instances to be separated and reflect nuclear heterogeneity as much as possible in the feature space. In the second stage, we freeze the weights of $f_{\theta } \left ( \cdot \right )$ and use it to extract instance-level features. These features are then concatenated to form bag-level representations, which are fed into the aggregator $g_{\tau }\left ( \cdot \right )$ along with corresponding bag-level labels for training.}
\label{fig:long}
\label{fig:onecol}
\end{figure*}

We undertake a reexamination of the limitations inherent in existing MIL frameworks\cite{ilse2018attention, lu2021data, li2021dual, shao2021transmil} and, leveraging the two aforementioned pathological priors, develop an innovative MIL framework, as illustrated in Figure 3. In order to eliminate noise from the background and optimize training efficiency, we use the OTSU algorithm to obtain foreground masks for the WSIs and generate patches at a specific magnification according to them. These patches are used to train a feature extractor with contrastive learning. For ease of exposition, we denote the set of patches obtained from the $i$-th slide as $X_{i} = \left \{ x_{i}^{1} , x_{i}^{2},\cdots, x_{i}^{n} \mid x_{i}^{k} \in \mathbb{R}^{H \times W \times 3} \right \}$, with corresponding labels $Y_{i}\in \mathbb{R}^C$. Given that the efficacy of contrastive learning is sensitive to the data augmentation scheme utilized during training\cite{chen2020simple, caron2020unsupervised, grill2020bootstrap}, we enhance the existing data augmentation strategy by incorporating random H\&E separation to improve instance-level embeddings with respect to nuclear heterogeneity. Subsequently, we pass the patches through the feature extractor on a per-bag basis, and concatenate them to obtain the bag-level embedding $E_i$. This process can be represented as:

\begin{align}
    \vec{e_i^k} = f_{\theta}\left (  x_{i}^{k}\right ) , & k=1, 2, \dots, n \\
    \vec{E_{i}} = & \left |  \right | _{k=1}^{n}\vec{e_i^k} 
\end{align}

\noindent where $\theta$ represents parameters of the extractor $f_{\theta}$, and $\left | \right |$ is concatenation operation. Note that during extraction, $\theta$ is frozen. Ultimately, $E_i$ serves as the input to train the aggregator $g_{\tau }$, as follows:
\begin{align}
    \hat{Y}_{i} = g_{\tau}\left (  E_i \right )  
\end{align}
\noindent where $\hat{Y_i}$ is the predicted label. In the following, we will present a detailed account of our data augmentation strategy as well as the specific structure of the aggregator.

\vspace{2ex}
\noindent \textbf{Contrastive Learning for Extractor.} \quad Due to the influence of staining agents, the color gamut of pathological images is significantly narrower than that of natural images. Therefore, features such as texture and shape are more critical than color. In most MIL methods, a ResNet pre-trained on ImageNet is directly transferred as the feature extractor. However, this often leads to poor instance discrimination in the feature space, as depicted in Figure 1. To address this issue and avoid introducing additional manual annotation, we employ contrastive learning, a self-supervised method, to train the feature extractor  $f_{\theta } \left ( \cdot \right )$. Among the state-of-the-art contrastive learning methods, SimSiam stands out for its ability to learn stable instance-level embeddings with even small batch size. 

In detail, all patches constitute a sample space $\Omega$ and are packed into batches. For each patch $x \in \Omega$, a pair of samples $\left(x_1, x_2 \right)$ is generated through random data augmentation. They serve as positive samples for each other, while all other samples in the same batch are negative samples for them. The pairs $\left(x_1, x_2 \right)$ are then fed into $f_{\theta}(\cdot)$ and a projection MLP to obtain their latent vectors $\left(z_1, z_2 \right)$, which are further fed into a prediction MLP to maximize their consistency. In particular, for H\&E pathological images, we add random H\&E separation to existing data augmentation scheme.

\vspace{2ex}
\noindent \textbf{Random H\&E Separation.} \quad To achieve images with single dye stained, we utilized the Vahadane method\cite{vahadane2016structure} for stain separation. For a given pathological image, the relative optical density matrix $V \in \mathbb{R}^{m \times n}$ can be expressed as a product of the stain color appearance matrix $W \in \mathbb{R}^{m \times r}$ and the stain density maps $H \in \mathbb{R}^{r \times n}$,  where $m$ is the number of channels, $r$ is the number of stains, and $n$ is the number of pixels, given by:
\begin{align}
    V=\log \frac{I_0}{I}=WH 
\end{align}
\noindent $I$ represents the matrix of RGB intensities, and $I_0$ is the illuminating light intensity (usually 255 for 8 bit images). Then, we can estimate W and H by solving the problem of sparse non-negative matrix factorization:
\begin{align}
\begin{aligned}
    \min_{W,H} \frac{1}{2}\left \| V-WH \right \|_{F}^2 & + \lambda \sum_{j=1}^{r}\left \| H\left ( j,: \right ) \right \|_1,      \\s.t. W,H \ge 0, & \left \| W\left ( :,j \right )  \right \|_2^2 = 1 
\end{aligned}
\end{align}

\noindent From the stain density maps $H$, we can derive the H channel and E channel images, denoted as $I_h$ and $I_e$, respectively:

\begin{align}
\begin{aligned}
    I_h = I_0 exp(- H[0, :]) \\
    I_e = I_0 exp(− H[1, :])
\end{aligned}
\end{align} 

\noindent In practical applications, it is imperative to preserve the color features of pathological images. To this end, we introduce a probabilistic parameter $p$, which stochastically converts patches to either their H-channel or E-channel images.

\begin{figure}[t]
\centering
   \includegraphics[width=\linewidth,trim=200 100 200 50,clip]{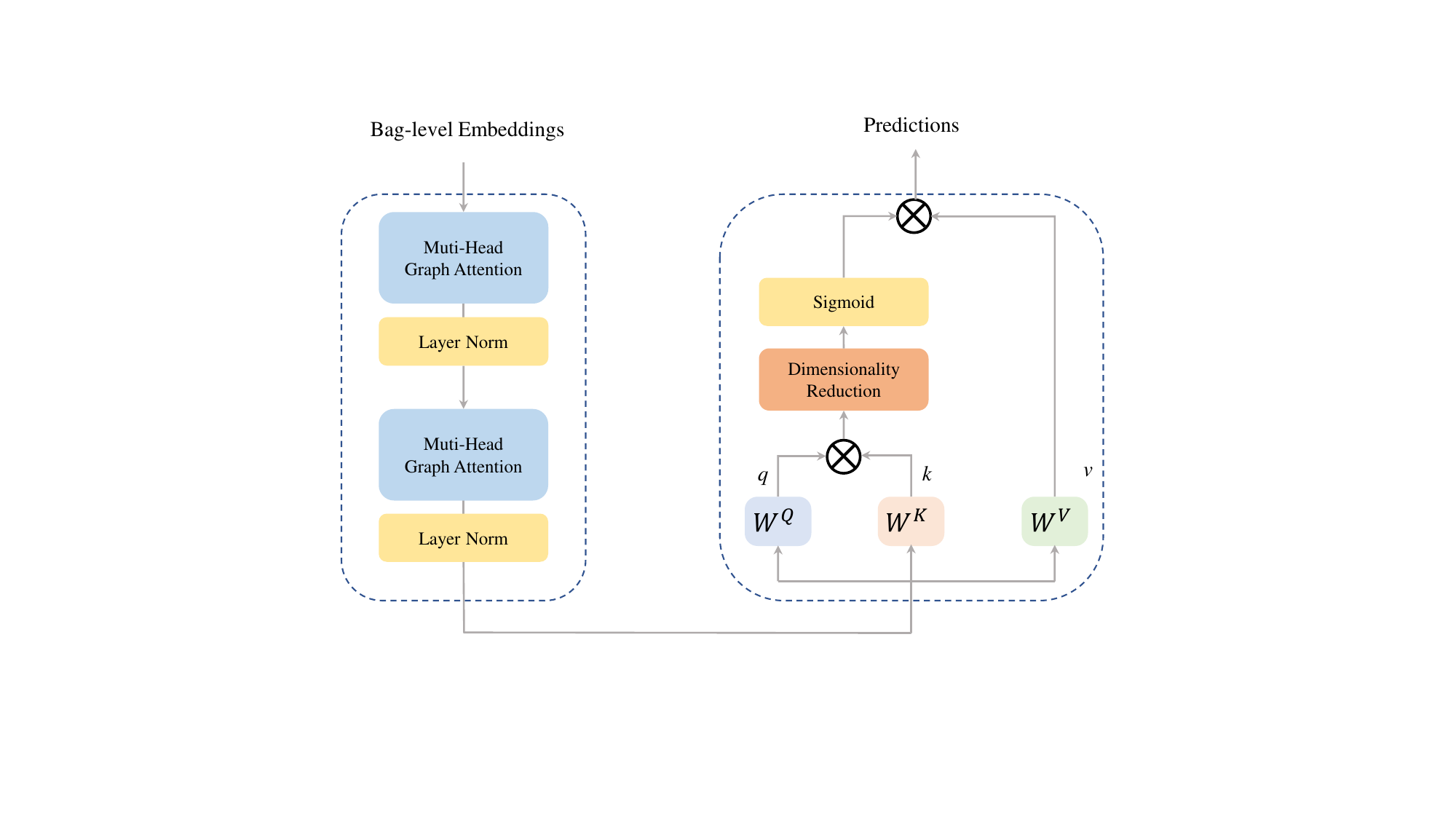}
   \caption{The Architecture of Graph Attention Aggregator. It consists of two graph-attention layers and one global-attention gate. Each instance-level representation is a node in the graph, and the relative positions of all instances in a bag are encoded as an adjacency matrix to limit the flow of mutual attention within the bag-level representation. The global-attention gate employs self-attention with dimensionality reduction to model the contribution of each instance, and the final prediction is obtained through a fully connected layer.}
\label{fig:long}
\label{fig:onecol}
\end{figure}

\vspace{2ex}
\noindent \textbf{Spatially-constrained aggregation network.} \quad In deep MIL models, attention mechanisms have been widely used in aggregators. Initially, it was recognized that although the bag-level embedding is a whole, it is actually composed of multiple instance-level embeddings. According to the formulation of MIL, these instance-level embeddings contribute differently to the final prediction. For example, in cancer detection, positive instances have larger weights in determining the final diagnosis of a slice, while negative instances in a negative slice have a more average impact on the result. Additionally, semantic information between instances should also be correlated. Therefore, a self-attention mechanism is used to model this, and it has been demonstrated to reduce the uncertainty of MIL\cite{shao2021transmil}. However, this approach has significant limitations, as there may not be intrinsic connections between every pair of instance-level embeddings. Analyzing the true distribution of lesions from the slices, only two instances that are spatially close are likely to have consistent attributes. Taking this into consideration, we propose an aggregator that employs graph attention\cite{velivckovic2017graph} to restrict the flow of mutual attention. In the final pooling stage, we design a global attention module that modifies the calculation method of self-attention to better fit the contributions of each instance, as shown in Figure 4.

In the graph attention module, the adjacency matrix $A$ of the graph attention layer is generated using the absolute positional indices of each instance:

\begin{align}
    A_{i,j} = \begin{cases}  0, & \text{ if } d_{i,j} > \sqrt{2}  \\  1, & \text{ if } 0 \le d_{i,j} \le \sqrt{2} \end{cases} 
\end{align}

\noindent where $d_{i,j}$ denotes the Euclidean distance between the coordinates of the $i$-th and $j$-th instances. Due to the non-negligible role of self-attention, the indices $i$ and $j$ may be equal. Following this, we leverage $A$ to compute masked attention, which solely assigns attention to the neighbor node set $N_i$ of instance $x_i$ (i.e., $x_j \in N_i$). The attention score between the node vector $\vec{e}_i$ of instance $x_i$ and the node vector $\vec{e}_j$ of its neighboring instance is then calculated as:
\begin{align}
    \alpha _{i,j} = \frac{exp(LeakyRelu(\vec{a}^T\left [ W\vec{e}_i\left |  \right |W\vec{e}_j   \right ]))  }{\sum _{k \in N_i}exp(LeakyReLu(\vec{a}^T\left [ W\vec{e}_i\left |  \right |W\vec{e}_k   \right ]))}  
\end{align}

\noindent In this equation, $W$ denotes a weight matrix that is responsible for performing a linear transformation on the input feature $\vec{e}_i$, while $\vec{a}$ represents a fully connected layer. Consequently, the output node vector $\vec{e^{'}}_i$ can be defined as follows:
\begin{align}
    \vec{e^{'}} _i = \sigma (\sum_{j \in N_i}\alpha_{i,j} W\vec{e}_j )  
\end{align}

\noindent In this context, $\sigma$ represents a non-linear activation function. Two successive graph attention layer are employed and during the final pooling stage, we utilize global-attention on the bag-level representation $E_i$ to derive weights for its spatial dimensions inspired by self-attention mechanism, which in turn model the contribution of the instance-level embeddings within the bag:
\begin{align}
    \vec{E_i^{'}} = sigmoid(\frac{\varphi( \vec{E_i} W_Q (\vec{E_i} W_K)^T) }{\sqrt{d_K} } )\vec{E_i}W_V  
\end{align}

\noindent The weight matrices $W_Q$, $W_K$ and $W_V$ are used to generate the query, key and value vectors, respectively. $\varphi$ means dimensionality reduction which we use to generate the spatial attention score, and average pooling was eventually chosen. $\vec{E_i^{'}}$ finally is projected into a low-dimensional space as the prediction for slide.

\section{Experiment}

\begin{table}[]
\begin{tabular}{ccccccl}
\hline
             & \multicolumn{3}{c}{Camelyon16}                                                                         & \multicolumn{3}{c}{TCGA-NSCLC}                                                                                                                \\ \cline{2-7} 
             & Accuracy                         & AUC                              & F1 score                         & Accuracy                                      & AUC                                           & F1 score                                      \\ \hline
mean pooling & 0.6667                           & 0.5283                           & 0.2712                           & 0.8140 $\pm$ 0.0169                           & 0.8323 $\pm$ 0.0231                           & 0.7999 $\pm$ 0.0210                           \\
max pooling  & 0.7597                           & 0.7635                           & 0.6804                           & 0.8345 $\pm$ 0.0062                           & 0.8634 $\pm$ 0.0263                           & 0.8223 $\pm$ 0.0067                           \\
MIL-Score    & 0.7752                           & 0.8120                           & 0.7929                           & 0.8563 $\pm$ 0.0036                           & 0.8834 $\pm$ 0.0111                           & 0.8550 $\pm$ 0.0077                           \\
MIL RNN      & 0.7829                           & 0.7834                           & 0.7021                           & 0.8575 $\pm$ 0.0072                           & 0.8754 $\pm$ 0.0093                           & 0.8557 $\pm$ 0.0070                           \\
ABMIL        & 0.8295                           & 0.8793                           & 0.7442                           & 0.8406 $\pm$ 0.0097                           & 0.8524 $\pm$ 0.0124                           & 0.8285 $\pm$ 0.0117                           \\
CLAM SB      & 0.7907                           & 0.7709                           & 0.6667                           & 0.8430 $\pm$ 0.0024                           & 0.8821 $\pm$ 0.0052                           & 0.8346 $\pm$ 0.0070                           \\
CLAM MB      & 0.8140                           & 0.8135                           & 0.7073                           & 0.8696 $\pm$ 0.0145                           & 0.8853 $\pm$ 0.0107                           & 0.8678 $\pm$ 0.0122                           \\
DSMIL        & 0.8217                           & 0.8527                           & 0.7356                           & 0.8853 $\pm$ 0.0084                           & 0.9004 $\pm$ 0.0252                           & \textbf{0.8855 $\pm$ 0.0099} \\
TransMIL     & 0.7984                           & 0.8189                           & 0.6977                           & 0.8804 $\pm$ 0.0205                           & 0.9042 $\pm$ 0.0310                           & 0.8733 $\pm$ 0.0209                           \\
Ours         & \textbf{0.8372} & \textbf{0.8876} & \textbf{0.7586} & \textbf{0.8877 $\pm$ 0.0109} & \textbf{0.9073 $\pm$ 0.0146} & 0.8785 $\pm$ 0.0157                           \\ \hline
\end{tabular}
\caption{The Results on Camelyon16 and TCGA-NSCLC Datasets}
\end{table}

In our experimentation, we evaluated our approach on two publicly available clinical pathology datasets: Camelyon16 and TCGA-NSCLC. These datasets offer a diverse range of MIL problems, spanning both balanced/unbalanced and single/multiple class scenarios. We conducted comparative experiments to assess the efficacy of our aggregator. Moreover, to corroborate the effectiveness of the individual components of our proposed framework, we carried out ablation studies.
\subsection{Dataset}
Camelyon16\cite{bejnordi2017diagnostic} is a publicly unbalanced dataset focused on differentiating between cancer and non-cancer cases for metastasis detection in breast cancer, which consists of 270 slides for training and 129 for testing. Following pre-processing with the OTSU algorithm, we acquired approximately 460,000 patches at 20× magnification, with an average of 1704 patches per slide.

TCGA-NSCLC\cite{tomczak2015review} includes two subtype projects, i.e., Lung Squamous Cell Carcinoma (LUSC) and Lung Adenocarcinoma (LUAD), for a total of 1034 diagnostic WSIs, including 527 LUAD slides and 507 LUSC slides. After pre-processing, the mean number of patches extracted per slide at 20× magnification is 3114.

\subsection{Experiment Setup and Evaluation Metrics.}
To obtain non-overlapping $256 \times 256$ patches, we employed the OTSU algorithm to generate foreground masks for all Whole Slide Images (WSIs). For the Camelyon16 dataset, we split the official training set into training and validation sets in an 8:2 ratio. The model was trained for 50 epochs on the training set and evaluated on the official test set to select the best-performing model on the validation set. For the TCGA-NSCLC dataset, we randomly divided all slides into 80\% training and 20\% validation, using four-fold cross-validation to assess model performance. We adopted accuracy, area under curve (AUC), and F1-score as evaluation metrics to measure the classification performance of model.

\subsection{Implementation Details}
To train the feature extractor, we utilized SimSiam\cite{chen2021exploring} and incorporated random H\&E separation in addition to random crop and color distortion. We employed Adam optimizer with an initial rate of 1e-4 and decayed the learning rate with the cosine decay schedule. The size of mini-batch was 256 and ResNet50 was selected as backbone. During MIL training, the feature of each patch is embedded in a 1024-dimensional vector by pre-trained extractor. We used Lookahead optimizer\cite{zhang2019lookahead} with a constant learning rate of 2e-4 and weight  decay of 1e-5. The mini-batch was 1.

\subsection{Baseline}
The baselines we chosed include deep models with traditional pooling operators such as mean-pooling, max-pooling and the current state-of-the-art embedding-level models, the attention gate based pooling operator ABMIL\cite{ilse2018attention}, non-local attention based pooling
operator DSMIL\cite{li2021dual}, single-attention-branch CLAM-SB\cite{lu2021data}, multi-attention-branch CLAM-MB\cite{lu2021data}, self-attention based aggregator TransMIL\cite{shao2021transmil}, and RNN based aggregation MIL-RNN\cite{campanella2019clinical}. Furthermore, we also evaluate the instance-level approach MIL-Score in our experiments.

\subsection{Slide-level Classification}

\begin{table*}[]
\begin{center}
\begin{tabular}{ccccccccc}
\hline
Method   &  & origin  & hematoxylin & eosin   &  & Accuracy & AUC    & F1 Score \\ \cline{3-5} \cline{7-9} 
ImageNet &  &         &             &         &  & 0.7984   & 0.7825 & 0.6977   \\
SimSiam  &  & $\surd$ &             &         &  & 0.8295   & 0.8859 & 0.7442   \\
SimSiam  &  & $\surd$ & $\surd$     &         &  & \textbf{0.8372}   & \textbf{0.8916} & 0.7529   \\
SimSiam  &  & $\surd$ & $\surd$     & $\surd$ &  & \textbf{0.8372}   & 0.8876 & \textbf{0.7586}   \\ \hline
\end{tabular}
\end{center}
\caption{The Influence of Different Feature Represetations on Camlyon16}
\end{table*}

Results of the cancer/non-cancer detection task on Camelyon16 and the subtypes classification task on TCGA-NSCLC are presented in Table 1. The experimental settings for other comparative methods are consistent with the official code. 

On the Camelyon16 dataset, only a small fraction of regions exhibit malignant growth. Moreover, the distribution of positive and negative slides manifests a remarkable degree of imbalance. All deep MIL models exhibit superior performance compared to traditional pooling operations. Furthermore, our proposed model outperforms the state-of-the-art ABMIL method in terms of accuracy, AUC, and F1 Score by 1.32\%, 1.24\%, and 1.44\%, respectively. Notably, the DSMIL method, which shares our approach of utilizing contrastive learning, also attains promising results. 

On TCGA-NSCLC, LUAD and LUSC exhibit significant differences in tissue structure, and the affected area accounts for over 80\% of the total tissue region. As there is no patch quantity imbalance between the two classes, instance-level MIL Score methods show great potential. Our proposed method achieves competitive performance compared to highly effective TransMIL and DSMIL models, with an improvement of 0.24\% and 0.31\% in accuracy and AUC, respectively. Notably, DSMIL outperforms our method in terms of F1 Score.

In addition, nuclear heterogeneity is of paramount importance in cancer detection, as cancerous cell nuclei exhibit distinct morphological differences from normal cell nuclei. As a result, our proposed method demonstrates a more significant improvement on Camelyon16, whereas it may not be the case on TCGA-NSCLC.

\subsection{Ablation Study}
Our model's primary contribution lies in the introduction of two pathological priors. With this in mind, we conducted a thorough ablation study on the pre-training strategy of the feature extractor, as well as the dimension decay method in the global attention gate of the aggregator. A series of comprehensive experiments conducted on the Camelyon16 dataset confirmed the effectiveness of these components.

\vspace{2ex}
\noindent\textbf{Pre-training Strategy.} \quad We utilized two feature extractor models: ResNet50 pre-trained on ImageNet and ResNet50 pre-trained using SimSiam. During the SimSiam training process, we employed three data augmentation schemes, including random cropping and color distortion, and additionally incorporated random H-separation and random H\&E-separation instead of random grayscale. We extracted features from the four strategies and used them as inputs to the aggregator to evaluate the effectiveness of our proposed random H\&E-separation. The experimental results presented in Table 2 demonstrate that the encoder pre-trained using SimSiam generally outperforms the one pre-trained on ImageNet, achieving an accuracy improvement of 3\%-4\%. This improvement is more pronounced in terms of AUC and F1 score. As the H-channel emphasizes the morphological characteristics of the nucleus, unlike simple random grayscale, and the E-channel only serves as a supplement, the performance of using random H-separation and random H\&E-separation is almost comparable.

\vspace{2ex}
\noindent\textbf{Dimensionality Reduction.} \quad In the aggregator structure, the global attention gate produces attention scores to weight the spatial dimensions and transform package-level embeddings into slice-level representations. We compared the effects of using max-pooling and average-pooling gates on the final performance. As shown in Table 3, the performance of average-pooling is consistently higher than that of max-pooling, with improvements of 1.55\%, 1.23\%, and 2.30\% in accuracy, AUC, and F1 score, respectively. 

\begin{table}[htbp]
\begin{center}
\begin{tabular}{cccc}
\hline
Reduction & Accuracy        & AUC             & F1 Score        \\ \cline{2-4} 
max.pool  & 0.8217          & 0.8793          & 0.7356          \\
avg.pool  & \textbf{0.8372} & \textbf{0.8876} & \textbf{0.7586} \\ \hline
\end{tabular}
\end{center}
\caption{Different Reduction Methods on Camelyon16}
\end{table}

%---------------------------------------------------------------------------------------------

\section{Conclusion}
In this paper, we propose a MIL framework based on two pathological priors, which has been shown to outperform previous methods on pathological datasets. Our key innovations are twofold. Firstly, we introduce a H\&E separation based data augmentation method that emphasizes nuclear heterogeneity and apply it to the pre-training of extractor. Secondly, we design an MIL aggregator based on the principle of positional similarity, which is highly interpretable. We use graph attention to calculate the mutual attention between only relevant patches and fit the weights of different instances through attention gate based on self-attention mechanism. 

In future research endeavors, the absence of comprehensive large-scale pathology standard databases accentuates the criticality of rational upstream pre-training methodologies. Within the context of MIL, self-supervised approaches warrant further investigation and comparison, including the utility of contrastive learning, autoencoder architectures, and generative adversarial networks, each with their unique advantages. Moreover, while multiscale information is increasingly valued within existing MIL paradigms, pre-processing complexities at different magnifications require due diligence, necessitating further discourse on effectively exploiting coarse and fine-grained features at a fixed magnification.

% \bibliographystyle{alpha}
% \bibliography{main}
\newcommand{\etalchar}[1]{$^{#1}$}

\end{CJK}
\end{document}